\documentclass[11pt, a4paper]{article}
\pdfoutput=1

\usepackage{amsmath}
\usepackage{amsfonts}
\usepackage{amssymb}
\usepackage{bm}
\usepackage{bbm}
\usepackage{verbatim}
\usepackage{dsfont}
\usepackage{booktabs}
\usepackage{slashed}
\usepackage{nicefrac}
\usepackage{mathtools}

\usepackage{epsfig}
\usepackage{color}
\usepackage[table]{xcolor}
\usepackage{graphicx}
\usepackage[font=small]{caption}
\usepackage[listofformat=empty,subrefformat=empty]{subfig}

\usepackage{float}
\usepackage{overpic}
\usepackage{tikz}

\usepackage{enumerate}
\usepackage{hhline}
\usepackage{multirow}

\usepackage[nosort]{cite}
\usepackage{xspace}
\usepackage{setspace}

\usepackage[pdftitle={Flavor Symmetries and Automatic Enhancement in the 6d Supergravity Swampland},
  pdfauthor={Mirjam Cvetic, Ling Lin, Andrew P. Turner},
  pdfsubject={Swampland, 6d supergravity, F-theory, symmetries},
  bookmarksopen, bookmarksnumbered, bookmarksopenlevel=2, colorlinks=false, linkcolor=blue, citecolor=blue, urlcolor=blue]{hyperref}

\usepackage[nameinlink,capitalize]{cleveref}

\makeatletter{}
\g@addto@macro\bfseries{\boldmath}
\makeatother

\newcommand*{\bbZ}{\ensuremath{\mathbb{Z}}}
\newcommand*{\bbR}{\ensuremath{\mathbb{R}}}
\newcommand*{\bbC}{\ensuremath{\mathbb{C}}}
\newcommand*{\bbP}{\ensuremath{\mathbb{P}}}
\newcommand*{\bbF}{\ensuremath{\mathbb{F}}}

\newcommand*{\cJ}{\ensuremath{\mathcal{J}}}
\newcommand*{\cN}{\ensuremath{\mathcal{N}}}
\newcommand*{\cZ}{\ensuremath{\mathcal{Z}}}

\newcommand*{\fkf}{\ensuremath{\mathfrak{f}}}
\newcommand*{\fkg}{\ensuremath{\mathfrak{g}}}
\newcommand*{\fkh}{\ensuremath{\mathfrak{h}}}

\DeclareMathOperator{\SO}{SO}
\DeclareMathOperator{\SU}{SU}

\DeclareMathOperator{\gE}{E}

\newcommand*{\fku}{\ensuremath{\mathfrak{u}}}
\newcommand*{\fksu}{\ensuremath{\mathfrak{su}}}
\newcommand*{\fkso}{\ensuremath{\mathfrak{so}}}
\newcommand*{\fksp}{\ensuremath{\mathfrak{sp}}}
\newcommand*{\fke}{\ensuremath{\mathfrak{e}}}

\DeclareMathOperator{\trace}{tr}

\newcommand*{\irrep}[2]{\ensuremath{\bm{\mathrm{R}}_{#1}^{(#2)}}}
\newcommand*{\charge}[2]{\ensuremath{q_{#1}^{(#2)}}}
\newcommand*{\birrep}{\ensuremath{\bm{\mathrm{R}}}}

\begin{document}

% format
\baselineskip=18pt  % a la harvmac
\numberwithin{equation}{section}  % make eq labels (sec.num)
\allowdisplaybreaks  % allow page breaks in displayed eqs

%%%%%%%%%%%%%%%%%%%%%%%%%%%%%%%%%%%%%%%%%%%
%%%        TITLE BEGINS HERE
%%%%%%%%%%%%%%%%%%%%%%%%%%%%%%%%%%%%%%%%%%%

\thispagestyle{empty}

\vspace*{-2cm}
\begin{flushright}
	\texttt{CERN-TH-2021-144} \\
	\texttt{UPR-1315-T}
\end{flushright}

%% title, authors, affiliation
\vspace*{0.6cm}
\begin{center}
{\LARGE{\textbf{Flavor Symmetries and Automatic Enhancement \vspace{.2cm}\\
in the 6d Supergravity Swampland}}} \\
 \vspace*{1.5cm}
Mirjam Cveti{\v c}$^{1,2,3,4}$, Ling Lin$^{4}$, Andrew P.~Turner$^{1}$\\

 \vspace*{1.0cm}
{\it ${}^1$ Department of Physics and Astronomy, University of Pennsylvania,\\
Philadelphia, PA 19104, USA\\
\it ${}^2$ Department of Mathematics, University of Pennsylvania,\\
Philadelphia, PA 19104, USA\\
${}^3$ Center for Applied Mathematics and Theoretical Physics,\\
University of Maribor, SI20000 Maribor, Slovenia\\
${}^4$CERN Theory Department, CH-1211 Geneva, Switzerland\\}

\vspace*{0.8cm}
\end{center}
\vspace*{.5cm}

\noindent
We argue for the quantum-gravitational inconsistency of certain 6d $\cN=(1,0)$ supergravity theories, whose anomaly-free gauge algebra $\fkg$ and hypermultiplet spectrum $M$ were observed in \cite{Raghuram:2020vxm} to be realizable only as part of a larger gauge sector $(\fkg' \supset \fkg, M' \supset M)$ in F-theory.
To detach any reference to a string theoretic method of construction, we utilize flavor symmetries to provide compelling reasons why the vast majority of such $(\fkg,M)$ theories are not compatible with quantum gravity constraints, and how the ``automatic enhancement'' to $(\fkg', M')$ remedies this.
In the first class of models, with $\fkg' = \fkg \oplus \fkh$, we show that there exists an unbroken flavor symmetry $\fkh$ acting on the matter $M$, which, if ungauged, would violate the No-Global-Symmetries Hypothesis.
This argument also applies to 1-form center symmetries, which govern the gauge group topology and massive states in representations different than those of massless states.
In a second class, we find that $\fkg$ is incompatible with the flavor symmetry of certain BPS strings that must exist by the Completeness Hypothesis.

\newpage
%%%%%%%%%%%%%%%%%%%%%%%%%%%%%%%%%%%%%%%%%%%
%%%           TITLE ENDS HERE
%%%%%%%%%%%%%%%%%%%%%%%%%%%%%%%%%%%%%%%%%%%
\setcounter{tocdepth}{2}
\tableofcontents
%\printindex

% \newpage

%%%%%%%%%%%%%%%%%%%%%%%%%%%%%%%%%%%%%%%%%%%
%%%        MAIN TEXT BEGINS HERE
%%%%%%%%%%%%%%%%%%%%%%%%%%%%%%%%%%%%%%%%%%%

\section{Introduction and Summary}

Physical features of string compactifications are highly constrained by geometric restrictions on the underlying compactification space.
One of the main insights from the Swampland program \cite{Vafa:2005ui} is the idea that, rather than limitations of string model building, these restrictions should reflect consistency conditions for coupling an effective quantum field theory to gravity (see \cite{Brennan:2017rbf,Palti:2019pca,vanBeest:2021lhn} for reviews).
To draw lessons about generic features of quantum gravity, it is therefore paramount that we understand the physical principles corresponding to these geometric restrictions.

With gauge symmetries being a central building block of the effective field theory framework, one important question to address is the landscape of possible symmetry groups.
In the context of effective supergravity theories, recent progress in this direction has been made in spacetime dimensions $d>6$ \cite{Adams:2010zy,Garcia-Etxebarria:2017crf,Kim:2019vuc,Cvetic:2020kuw,Montero:2020icj,Hamada:2021bbz}, by finding physical consistency conditions that lead to the same restrictions as in known string compactifications, such as bounds on the total gauge rank as well as the allowed gauge algebras and their global structures.

In 6d, gauge dynamics of $\cN = (1, 0)$ supergravity theories is already heavily constrained by the cancellation of chiral anomalies.
Nevertheless, the landscape of string-derived models, in particular F-theory constructions \cite{Vafa:1996xn,Morrison:1996na,Morrison:1996pp}\footnote{See \cite{Weigand:2018rez,Cvetic:2018bni} for recent reviews on F-theory.}, appears to be much smaller compared to the vast set of anomaly-free field theories \cite{Kumar:2009us,Kumar:2009ae,Kumar:2010ru}.
Utilizing Swampland principles \cite{Seiberg:2011dr,Monnier:2017oqd,Monnier:2018nfs,Kim:2019vuc,Lee:2019skh,Angelantonj:2020pyr,Apruzzi:2020zot,Tarazi:2021duw,Cheng:2021zjh}, we have since learned that, indeed, most of such field theories are inconsistent when coupled to quantum gravity.
However, our knowledge about the boundary between the Swampland and Landscape in the context of 6d supergravity theories is still blurred when it comes to the precise set of consistent gauge symmetries.

A recent proposal \cite{Raghuram:2020vxm}, dubbed the ``Automatic Enhancement Conjecture'', attempts to sharpen this boundary considerably.
This is based on the observation that, in engineering certain gauge algebras $\fkg$ with matter $M$ via F-theory, the underlying geometry forces the appearance of a ``larger'' symmetry $\fkg' \supset \fkg$, whose matter spectrum $M' \supset M$ does not allow for a supersymmetric Higgsing of $\fkg'$ to $\fkg$.
In other words, it appears that F-theory can only realize $(\fkg,M)$ as a subsector of a larger gauge symmetry, even though all gauge and gravitational anomalies are cancelled.\footnote{More precisely, any 6d gauge symmetry has an associated anomaly coefficient $b_\fkg$, which is part of the defining data of the supergravity model. See \cref{app:anomaly_cancellation} for more details.}
This motivated the conjecture \cite{Raghuram:2020vxm} that, for such a pair of theories, $(\fkg' \supset \fkg, M' \supset M)$, where the larger $(\fkg',M')$ can be constructed in F-theory, the smaller $(\fkg,M)$ theory belongs to the Swampland---the set of apparently consistent effective gravitational theories which cannot be UV-completed.
This formulation makes explicit reference to an F-theoretic method of construction, as the enhanced gauge group $\fkg'$ was found through explicit geometric engineering.
The geometric motivations obscure the physical ``origin'' of $(\fkg',M')$ apart from the non-existence of a supersymmetric Higgs transition $(\fkg', M') \rightarrow (\fkg, M)$, and, in particular, do not explain why the $(\fkg, M)$ theory is inconsistent by itself.

The goal of this work is to shed light on the physical arguments underlying the Automatic Enhancement Conjecture.
While our analysis reveals no single, unifying principle, two sets of arguments---related to flavor symmetries---emerge that lend credence to the conjecture in different types of automatic enhancement observed in \cite{Raghuram:2020vxm}.

The first type of enhancement we seek to explain are models where $\fkg' = \fkg \oplus \fkh$, with the matter $M' = \bigoplus_{r} x'_{r} \times (\irrep{\fkg}{r}, \irrep{\fkh}{r})$ (for $\irrep{\fkf}{r}$ representations of $\fkf = \fkg, \fkh$) being just a rearrangement of the original matter $M = \bigoplus_{r} (x'_{r} \dim \irrep{\fkh}{r}) \times  \irrep{\fkg}{r}$ into bi-charged representations.
In these instances, we show that $\fkh$ has the interpretation of the subalgebra of the classical flavor symmetry $\hat{\fkh}$ of $(\fkg, M)$ for which the flavor--gauge, flavor--flavor, and flavor--gravitational anomalies can be cancelled by a generalized Green--Schwarz (GS) mechanism \cite{Green:1984bx,Sagnotti:1992qw} with the tensor multiplets present in the theory.
The characterization of $\fkh$ as a subalgebra of the flavor symmetry implies in particular that there is no Higgsing from $(\fkg', M')$ to $(\fkg, M)$, since an honest flavor symmetry of $\fkg$ only charges matter that is already charged under $\fkg$, which would then also break $\fkg$ if given a vacuum expectation value (vev).
The automatic enhancement to $(\fkg' = \fkg \oplus \fkh, M' \cong M)$ is then the result of an equivalent formulation of the conjecture that needs no reference to any string theory construction:
\begin{center}
\emph{If the flavor symmetry of $(\fkg, M)$ has a subalgebra $\fkh$ whose anomalies can be cancelled by an appropriate GS term, then, in a consistent 6d supergravity model, $\fkh$ must also be gauged.}
\end{center}
In particular, it implies that any potential sources of symmetry breaking in the UV are entirely captured by anomalies in the low-energy effective gauge theory.
Therefore, the absence of these anomalies implies that the $(\fkg, M)$ theory is inconsistent as an effective description of quantum gravity, because it has an unbroken global symmetry.
To conform with the No-Global-Symmetries Hypothesis \cite{Banks:2010zn,ArkaniHamed:2006dz,Polchinski:2003bq}, this symmetry must hence be gauged in quantum gravity.
In \cref{sec:flavor_anomalies}, we will discuss this interpretation in more detail, and provide detailed examples.

Moreover, we also relate the global structure of the gauge group in these cases, which arises in F-theory models ``automatically'' in form of the Mordell--Weil group \cite{Aspinwall:1998xj,Mayrhofer:2014opa,Cvetic:2017epq}, to an automatic enhancement of 1-form center symmetries \cite{Gaiotto:2014kfa}, which in 6d $\cN=(1,0)$ theories have an anomaly \cite{Apruzzi:2020zot} that obstruct their gauging.
As we point out in \cref{sec:1-form_anomalies}, the cases in \cite{Raghuram:2020vxm,Morrison:2021wuv} where the gauge \emph{group} associated to the enhanced algebra $\fkg'$ is automatically non-simply connected have no such obstruction.
Hence, the automatic gauging of the 1-form symmetry, leading to the non-simply connected gauge group, can be understood equivalently as an incarnation of the No-Global-Symmetries Hypothesis, now for generalized global symmetries.
In \cref{sec:comments}, we further comment on the relationship of the aforementioned obstructions to the validity of the ``Massless Charge Sufficiency Conjecture'' \cite{Morrison:2021wuv} in 6d supergravity theories.

While these arguments provide a consistent interpretation of the Automatic Enhancement Conjecture in the larger web of quantum gravity constraints, this bottom-up perspective does not provide the means to rule out UV-breaking mechanisms not captured by anomalies; the observed enhancement is only a post-factum confirmation of the absence of such a breaking.
Hence, this reformulation, while revealing the intimate connection to the No-Global-Symmetries Hypothesis, does not constitute a definitive proof of the conjecture.
Moreover, there are enhancements $(\fkg, M) \to (\fkg',M')$ found in \cite{Raghuram:2020vxm} that do not have an interpretation as gauging a flavor symmetry of $(\fkg, M)$, and, thus, would require a different set of arguments to explain.

As we will argue, the key objects to formulating such arguments are BPS strings.
However, because the Automatic Enhancement Conjecture places a \emph{lower} bound on the allowed gauge symmetry, the methods of \cite{Kim:2019vuc,Lee:2019skh,Cheng:2021zjh} to utilize the so-called supergravity string---a type of BPS string present in all 6d $\cN=(1,0)$ supergravity theories---are not immediately applicable, as these place upper bounds.
Instead, we find that one must inspect strings specific to the model, whose worldsheet consistency then provides constraints equivalent to the Automatic Enhancement Conjecture.
In 6d $\cN = (1, 0)$ theories, such strings must exist by virtue of the Completeness Hypothesis \cite{Polchinski:2003bq} as charged objects of 2-form fields in tensor multiplets, which themselves are required to cancel gauge anomalies via the Green--Schwarz mechanism.
Then, the consistency of the worldsheet CFT poses conditions on the spacetime gauge dynamics beyond effective field theory considerations.
In particular, from the worldsheet perspective, it is now the spacetime symmetry $\fkg$ that plays the role of a global, or flavor, symmetry.
In this case, the specific type of strings (with charge $b$) present in the supergravity theory can constrain the symmetry algebra $\fkg$, even if it is consistent with gauge anomalies in the 6d bulk.
From this perspective, the enhancement $\fkg \to \fkg \oplus \fkh$ observed in F-theory constructions reflects the geometry's ``awareness'' of these constraints.
It avoids the string of charge $b$ with an inconsistent flavor symmetry $\fkg$ by ``pairing'' it with a spacetime gauge symmetry $\fkh$, i.e., only allows instanton strings of $\fkh$ with compatible flavor symmetry to have charge $b$.
In \cref{sec:strings}, we exemplify this principle in two classes of models, where the constraining strings are strings of 6d $\cN = (1, 0)$ superconformal field theories (SCFTs) with well-known flavor symmetries.
In \cref{sec:discussion}, we will discuss some open questions about the Automatic Enhancement Conjecture that will need further study of BPS strings of 6d supergravity theories to elucidate.

\section{Automatic enhancement as gauging of flavor symmetries}
\label{sec:flavor_anomalies}

In this section, we focus on the bottom-up interpretation of the Automatic Enhancement Conjecture as quantum gravity avoiding unbroken global symmetries.
First, we verify for enhancements $(\fkg, M)$ to $(\fkg' = \fkg \oplus \fkh, M' \cong M)$, with no hypermultiplets charged solely under $\fkh$, that $\fkh$ is the subalgebra of the flavor symmetry $\hat\fkh$ of $(\fkg,M)$ for which the flavor--gauge, flavor--flavor (or 't~Hooft), and flavor--gravitational anomalies can be cancelled by a generalized Green--Schwarz mechanism.
We then use the same perspective to explain the global gauge group topology, as quantum gravity avoiding an unbroken 1-form global symmetry.

\subsection{Review of 6d gauge anomalies}
\label{app:anomaly_cancellation}

To begin with, we review gauge and gravitational anomalies of 6d $\cN = (1, 0)$ gauge and supergravity theories, which also explains our notation.
For a pedagogical discussion, see, e.g., \cite{Taylor:2011wt}. We are not concerned here with global anomalies; for more on such anomalies, see \cite{BershadskyVafaAnom6D,SuzukiTachikawaAnom6D,KumarMorrisonTaylor6DSUGRA}.

Six-dimensional $\cN = (1, 0)$ supergravity theories famously have an associated signature-$(1,T)$ unimodular lattice $\Gamma$ with pairing $\Omega$ that corresponds to the charge lattice of one self-dual and $T$ anti-self-dual 2-form fields (or, dually, with the BPS strings charged under these 2-forms) in the supergravity and tensor multiplets, respectively.
The vevs of the scalars in these multiplets are parametrized by a vector $\cJ \in \Gamma \otimes \bbR$ of unit length under $\Omega$, which in turn defines a \emph{positivity cone} $\Gamma^+$, given by the requirement $\Omega(v, \cJ) \equiv v \cdot \cJ \ge 0$ for any $v \in \Gamma^+$.
There is a model-specific vector $a \in \Gamma$, called the \emph{gravitational anomaly coefficient}.
To complete the data for a $(1,0)$ supergravity theory, we have to specify the gauge algebra,
    \begin{equation}
        \fkg = \bigoplus_i \fkg_i \oplus \bigoplus_\alpha \fku(1)_\alpha\,,
    \end{equation}
and hypermultiplet spectrum
\begin{equation}\label{eq:hyper_total_rep}
    M = \bigoplus_r x_r \times \irrep{\fkg}{r} \equiv \bigoplus_r x_r \times (\irrep{\fkg_1}{r}, \dotsc, \irrep{\fkg_i}{r}, \dotsc)_{\charge{1}{r}, \dotsc, \charge{\alpha}{r}, \dotsc}\,.
\end{equation}
Here, the irreducible representations (irreps) $\irrep{\fkg}{r}$ with multiplicity $x_r$ are distinguished by irreps $\irrep{\fkg_i}{r}$ under each non-abelian gauge factor $\fkg_i$, and by charges $q_\alpha^{(r)}$ under each abelian $\fku(1)_\alpha$ factor.\footnote{Note that a hypermultiplet we refer to as being in a representation $\birrep$ actually contains field transforming in both $\birrep$ and $\overline{\birrep}$, because the hypermultiplets live in a quaternionic-K\"{a}hler manifold and thus can only be charged under quaternionic representations of the gauge group. When the representation $\birrep$ is itself quaternionic, we may form half-hypermultiplets that have half the degrees of freedom of a full hypermultiplet.\label{footnote_irreps}}

We associate to each non-abelian factor $\fkg_i$ an anomaly coefficient $b_i \in \Gamma^+$, and to each \emph{pair} of abelian gauge factors an anomaly coefficient $b_{\alpha \beta} = b_{\beta \alpha}$ that satisfies $\frac12 b_{\alpha \alpha}, b_{\alpha \beta} \in \Gamma$, and for which $\cJ \cdot b_{\alpha \beta}$ is a positive definite matrix.\footnote{Note that the integrality conditions, i.e., being elements of $\Gamma$ rather than $\Gamma \otimes \bbR$, are related to a generalized Completeness Hypothesis, see \cite{Monnier:2017oqd}.}
A tensor multiplet with charge $b$ that is an anomaly coefficient of a gauge factor $\fkg_i$ is said to be ``paired'' with the corresponding vector multiplet \cite{Morrison:2016djb}.
Equivalently, this means that the 1-instanton configuration of $\fkg_i$, which is string-like in 6d, has charge $b$.

These anomaly coefficients encode the contributions to the total anomaly polynomial $I_8^\text{total} = I_8^\text{1-loop} + I_8^\text{GS}$ from a generalized version of the Green--Schwarz mechanism \cite{Green:1984bx,Sagnotti:1992qw} via the inclusion of tree-level diagrams mediated by the (anti-)self-dual 2-form fields:
\begin{equation}
    I_8^\text{GS} = \frac{1}{32} \Omega(X_4, X_4) \equiv \frac{1}{32} X_4 \cdot X_4 \,,
\end{equation}
with
\begin{equation}\label{eq:GS-term_general}
    X_4 = \frac12 a \trace R^2 + 2 \sum_i \frac{b_i}{\lambda_i} \trace F_i^2 + 2 \sum_{\alpha,\beta} b_{\alpha \beta} F_\alpha F_\beta \,,
\end{equation}
where $F_\alpha$, $F_i$ are the field strengths of the (non-)abelian gauge symmetries, $R$ is the spacetime curvature 2-form, and $\trace$ denotes trace in the fundamental representation.
The $\lambda_i$ are normalization constants associated with the simple non-abelian Lie algebras, given by $\lambda_i = 2 c_i^\vee / A_{\fkg_i}$ with $c_i^\vee$ the dual Coxeter number.
Meanwhile, the 1-loop part $I_8^\text{1-loop}$ receives contributions from the gravity, tensor, vector, and hypermultiplets.
Their explicit expressions, as well as the resulting conditions for anomaly cancellation, $I_8^\text{total} = 0$, and the values for $\lambda_i$ are summarized in \cref{app:anomaly_conditions}.

\subsection{Cancellation of flavor anomalies}
\label{sec:approach}

From the bottom-up perspective, it is natural to think about a supergravity theory as coupling an effective 6d $\cN = (1,0)$ field theory with gauge algebra $\fkg$ and massless matter $M$ to gravity in a specific way.
For the $(\fkg, M)$ gauge theory, there is an associated classical flavor symmetry $\hat{\fkh}$ that rotates $M$.\footnote{Associated to the decomposition \labelcref{eq:hyper_total_rep} of the hypermultiplets $M$, $\hat{\fkh}$ contains a $\fku(x_r)$ factor rotating a complex irrep $\irrep{\fkg}{r}$ with multiplicity $x_r$, an $\fksp(x_r)$ rotating a real irrep $\irrep{\fkg}{r}$ (with $\fksp(1) \cong \fksu(2)$), and an $\fkso(2x_r)$ rotating a quaternionic irrep $\irrep{\fkg}{r}$; note that in the last case, $x_r$ can be half-integer, see footnote \ref{footnote_irreps}.}
This re-organizes the matter \labelcref{eq:hyper_total_rep} as
\begin{equation}\label{eq:reps_under_flavor}
    M = \bigoplus_r x_r \times \irrep{\fkg}{r} \cong \bigoplus_r \irrep{\fkg \oplus \hat{\fkh}}{r} = \bigoplus_r (\irrep{\fkg}{r}, \irrep{\hat{\fkh}}{r})\,,
\end{equation}
with $x_r = \dim \irrep{\hat{\fkh}}{r}$. We are interested in the cancellation of flavor anomalies, which allows for gauging a part $\fkh \subset \hat{\fkh}$ of the flavor symmetry, i.e., all flavor--gauge, flavor--flavor, and flavor--gravitational anomalies associated with $\fkh$.
To cancel the 1-loop contributions, one must introduce appropriate Green--Schwarz terms, which structurally take the same form as in \cref{eq:GS-term_general}, but now include the field strengths of $\fkh$.
In particular, this requires the introduction of anomaly coefficients $b_I$, $b_{\alpha \hat{\beta}}$, and $b_{\hat{\alpha} \hat{\beta}}$  for the flavor factors $\fkh = \bigoplus_I \fkh_I \oplus \bigoplus_{\hat{\alpha}} \fku(1)_{\hat{\alpha}}$; in terms of the effective field theory action, this requires a coupling of the 2-form field dual to $b_I$ to the instanton density of $\fkh$.

In the context of 6d superconformal field theories (SCFTs), such a gauging procedure (of non-abelian flavor factors) can be used to construct new SCFTs out of existing ones.
However, this always introduces a new anomaly coefficient $b_I$ that is linearly independent of those already present.
In other words, the lattice of string/tensor charges is extended in these gaugings.
On the other hand, the Automatic Enhancement Conjecture describes a change of gauge symmetry without altering the tensors.
In other words, the required Green--Schwarz terms must be supplied by the existing tensors of the supergravity model prior to enhancement.
Moreover, we demand that this is possible without the introduction of additional charged hypermultiplets that are charged under $\fkh$.
This retains the interpretation of $\fkh$ as ``just the flavor symmetry'' of the gauge sector $(\fkg, M)$, and, furthermore, prevents $\fkh$ from being Higgsed without also breaking $\fkg$.
In short, we would like to determine the subalgebra $\fkh \subset \hat{\fkh}$ for which all 1-loop anomalies induced by the hypermultiplet spectrum \labelcref{eq:reps_under_flavor} are cancelled by the $\fkh$-vector multiplets and Green--Schwarz terms built out of tensors already present in the $(\fkg, M)$ theory.

If this is possible, then, from a bottom-up perspective, there are no obstructions to turning on a ``background gauge field'' for the flavor symmetry $\fkh$---after all, the anomalies precisely measure such obstructions that are detectable in the effective field theory description.
However, in a gravitational theory, there are no background gauge fields; in 6d $\cN=(1,0)$ theories, this would require the gauge coupling $g_I^2 \propto (\cJ \cdot b_I)^{-1}$ to vanish, which gives an infinite tension $\cJ \cdot b_I$ to the string dual to $b_I$.
Therefore, the background field for $\fkh$ is allowed to fluctuate (note that this requires also the cancellation of what would be 't~Hooft anomalies for $\fkh$ if treated as a global symmetry), and, in particular, gauges $\fkh$, thus avoiding an unobstructed global symmetry.
We have verified for all relevant examples discussed in \cite{Raghuram:2020vxm} that the observed automatic enhancement is precisely such an automatic gauging of an unobstructed flavor symmetry $\fkh$.

To perform this verification, one would, ideally, consider every possible subalgebra embedding $\fkh = \bigoplus_I \fkh_I \oplus \bigoplus_{\hat{\alpha}} \fku(1)_{\hat{\alpha}} \subset \hat{\fkh}$ and consider, for each of these cases, if the associated flavor--gauge, flavor--flavor, and flavor--gravitational anomalies can be cancelled by assigning anomaly coefficients $b_I, b_{\alpha \hat{\beta}}, b_{\hat{\alpha} \hat{\beta}} \in \Gamma$.
However, this approach can become quite computationally demanding when the flavor symmetry is large and thus has a very large number of subalgebras. We propose here an alternate approach that allows for a more direct computation of the maximal gauge-able subalgebra of $\hat{\fkh}$.

We first consider gauging an arbitrary $\fku(1)$ subalgebra of $\hat{\fkh}$, and then consider the constraints on the charges of the resulting matter spectrum following from the 6d anomaly cancellation conditions \labelcref{eq:6dAC}.
In practice, this is done by augmenting the gauge algebra $\fkg$ (with the anomaly coefficients $b_i, b_{\alpha \beta}$) with an additional $\fku(1)$ factor and considering all possible assignments of $\fku(1)$ charges for irreps $\irrep{\fkg}{r}$ of the original matter $M$ and anomaly coefficient $b_{\fku(1)}$ that solve the anomaly cancellation conditions.

Crucially, in solving for anomaly-consistent spectra, we allow for a potentially \emph{negative} number of hypermultiplets charged only under the new $\fku(1)$ gauge factor.
The presence of a negative number of such hypermultiplets signals that the newly-introduced $\fku(1)$ must be embedded in a larger non-abelian subalgebra of $\hat{\fkh}$ in order to be consistent.

More precisely, imagine the $\fku(1)$ arising from an adjoint Higgsing of $\fkh$, in which hypermultiplets charged solely under $\fkh$ generically decompose into charged singlets.
These are counted against those hypermultiplets that are ``eaten'' by the vectors in the Higgs transition.
A formally negative number of such charged singlets in the Higgsed phase therefore signals that there were not enough hypermultiplets to begin with to facilitate this Higgsing of $\fkh$.
The exact (negative) number of charged singlets in the spectrum restricts the simple non-abelian subalgebras $\fkh$ of $\hat{\fkh}$ into which the $\fku(1)$ factor can consistently be embedded.
One can then verify whether this candidate algebra is anomaly-free with the anomaly coefficient $b_\fkh \propto b_{\fku(1)}$.

When solving the anomaly cancellation conditions for $\fkg \oplus \fku(1)$, we allow for the possibility that each hypermultiplet in a given representation $\irrep{\fkg}{r}$ has a different $\fku(1)$ charge, to accommodate any possible embedding into a larger algebra. However, we only consider (non-positive multiplicities of) singlets of charge $1$ and $2$ under the additional $\fku(1)$ factor. The reason for this is that, although adjoint Higgsings of simple non-abelian factors can result in singlets of other charges, higher-charge representations can always be ``exchanged'' via anomaly equivalences until only charges $1$ and $2$ remain, and the number of uncharged singlets will always increase in such an exchange. Specifically, the following anomaly equivalence holds \cite{TaylorTurnerGeneric}:
    \begin{equation}
        \label{eq:u1-equivalence}
        \bm{1}_q
        \longleftrightarrow
        -\frac{q^2 \left(q^2 - 4\right)}{3} \times \bm{1}_1 +
        \frac{q^2 \left(q^2 - 1\right)}{12} \times \bm{1}_2 +
        \left(\frac{q^4}{4} - \frac{5 q^2}{4} + 1\right) \times \bm{1}_0\,.
    \end{equation}
We see that a single charge-$q$ singlet is equivalent under anomalies to a negative number of charge-$1$ singlets along with  positive numbers of charge-$2$ and charge-$0$ singlets. As we are only using the charged singlets as a computational tool to identify that the $\fku(1)$ factor is embedded in a gauged non-abelian algebra, this allows us to reduce the scope of the computation. Furthermore, we can reduce the list of possible charges of other hypermultiplets by considering analogous anomaly equivalences for $\fku(1)$-charged non-abelian representations other than singlets. Including higher $\fku(1)$-charged representations will always reduce the number of uncharged singlets; because each model $(\fkg, M)$ we consider has only a finite number of uncharged singlets, there is a limit to magnitude of the $\fku(1)$ charges under the new $\fku(1)$ factor before the number of uncharged singlets would be forced to be negative.

We stress that this procedure is merely a computational simplification to find the gauge-able subalgebra $\fkh$.
We have not investigated potential exceptions to this algorithm, since in all cases we utilize it, we find a consistent subalgebra $\fkh \subset \hat{\fkh}$, which furthermore agrees with the instances of automatic enhancement observed \cite{Raghuram:2020vxm}.
As we will see in \cref{sec:su3_example}, the algorithm might lead to an intermediate $\fku(1)$ spectrum that does not embed into the full flavor $\hat{\fkh}$, but, in addition, also to a spectrum compatible with the Cartan charges of a subalgebra $\fkh \subset \hat{\fkh}$ that agrees with the automatic enhancement.
Since the alternative spectrum has a negative number of some singlets, it is inconsistent by itself anyways, so does not pose a violation of the Automatic Enhancement Conjecture per se.\footnote{We have also not found a consistent non-abelian algebra into which such spectra embed, such that all negative number of singlets are ``used up'', thus lending further evidence to the efficacy of the algorithm.}

In all examples we have examined, we only find one unique subalgebra $\fkh \subset \hat{\fkh}$, which reflects the fact that in all models studied in \cite{Raghuram:2020vxm}, there is a unique automatic enhancement.
However, it is in principle possible that there exist multiple (potentially identical) subalgebras $\fkh_\ell$ of $\hat{\fkh}$ that can be consistently gauged with different anomaly coefficients $b_\ell$, all of which would be detectable by the algorithm above.
By definition, there would not be any supersymmetric Higgs transitions that flow from one case to another, so, from our bottom-up perspective, the gauging of any $\fkh_\ell$ with $b_\ell$ would be a consistent model.
If such cases exist, then this must happen for $T \ge 1$, in which case it would be interesting to investigate, from an F-theory perspective, how the multiple allowed enhancements may be realized.
Note that there can be additional consistency constraints for $T \ge 1$ such as those that we discuss in \cref{sec:strings}, which could potentially exclude these cases.

\subsection{Examples}
\subsubsection{Large \texorpdfstring{$\SU(N)$}{SU(N)} gauge group with \texorpdfstring{$T = 0$}{T=0} and \texorpdfstring{$b = 1$}{b=1}}

To illustrate the procedure, we will demonstrate how it is carried out for the enhancements of large $\fksu(N)$ gauge algebras with $T = 0$ and $b = 1$; these appear in Section~4.3 of \cite{Raghuram:2020vxm}. We will begin with the case $N = 22$, where the flavor symmetry is small enough that we can carry out the exhaustive computation mentioned in \cref{sec:approach}, before addressing the same example and all subsequent ones using the shortcut approach described there.

For the gauge symmetry $\fksu(22)$ and $T = 0$, we consider the hypermultiplet spectrum
    \begin{equation}
        \label{eq:spectrum_su22}
        x_{\bm{1}} = 19\,, \quad x_{\bm{22}} = 2\,, \quad x_{\bm{231}} = 3\,, \quad x_{\bm{483}} = 0\,,
    \end{equation}
which solves the anomaly cancellation conditions with anomaly coefficient $b = 1$. Here, $x_{\birrep}$ denotes the multiplicity of the representation $\birrep$. The full flavor symmetry of this model is $\hat{\fkh} = \fku(2) \oplus \fku(3)$, associated with the $2$ fundamental and $3$ anti-symmetric representations. We thus want to find the subalgebras of $\fku(2) \oplus \fku(3)$ which can be gauged.

First, we note that the $\fksu(2) \subset \fku(2)$ associated with the fundamental flavors is non-anomalous; to see this, we look for solutions to the anomaly cancellation conditions \labelcref{eq:6dAC} for an $\fksu(22) \oplus \fksu(2)$ algebra with $T = 0$ and $b_{\fksu(22)} = 1$ that only have $\fksu(22)$ fundamentals charged under the additional $\fksu(2)$ factor. The unique solution is
    \begin{equation}
        \label{eq:su22su2fund}
        x_{(\bm{1}, \bm{1})} = 22\,, \quad x_{(\bm{22}, \bm{1})} = 0\,, \quad x_{(\bm{231}, \bm{1})} = 3\,, \quad x_{(\bm{483}, \bm{1})} = 0\,, \quad x_{(\bm{22}, \bm{2})} = 1\,,
    \end{equation}
with $b_{\fksu(2)} = 1$. Thus, the $\fksu(2)$ fundamental flavor symmetry can be gauged.

Next, we can consider gauging a subalgebra of the $\fksu(3)$ anti-symmetric flavor symmetry. Similarly to above, we can look for solutions to the anomaly cancellation conditions in which only the $\fksu(22)$ anti-symmetric representations are allowed to have non-trivial charge under the additional gauge factor; we find that there are no valid solutions for any subalgebra of the $\fksu(3)$. To see how this occurs, consider the case of $\fksu(22) \oplus \fksu(3)$. The solutions to the anomaly cancellation equations take the form
    \begin{equation}
        \begin{aligned}
            x_{(\bm{1}, \bm{1})} &= 5 b_{\fksu(3)}^2 - \frac{687}{20} b_{\fksu(3)} + 19\,, \\
            x_{(\bm{22}, \bm{1})} &= 2\,, \\
            x_{(\bm{231}, \bm{1})} &= 3 - \frac{3}{20} b_{\fksu(3)}\,, \\
            x_{(\bm{1}, \bm{3})} &= b_{\fksu(3)} \left(\frac{309}{20} - 3 b_{\fksu(3)}\right)\,, \\
            x_{(\bm{483}, \bm{1})} &= 0\,, \\
            x_{(\bm{1}, \bm{8})} &= \frac{1}{2} (b_{\fksu(3)} - 1) (b_{\fksu(3)} - 1)\,, \\
            x_{(\bm{231}, \bm{3})} &= \frac{b_{\fksu(3)}}{20}\,.
        \end{aligned}
    \end{equation}
Requiring that this $\fksu(3)$ is subgroup of the flavor $\fku(3)$ of the $\fksu(22)$ anti-symmetrics sets $x_{(\bm{231}, \bm{3})} = 1$ and $x_{(\bm{1}, \bm{3})} = 0$, which cannot be solved simultaneously; thus, the $\fksu(3)$ cannot be gauged.
Similar results hold for all subgroups of the $\fksu(3)$. We then proceed to check anomalies for the two $\fku(1)$ factors; the fundamental $\bm{22}$ is charged only under the $\fku(1) \subset \fku(2)$, while the anti-symmetric $\bm{231}$ is charged only under the second $\fku(1) \subset \fku(3)$. Again, we find that the anomaly conditions cannot be solved without the introduction of further charged hypermultiplets. To rule out potential linear combinations of Cartan $\fku(1)$s that can be gauged, we must then also consider linear combinations of $\fku(1)$ charges for the bifundamental and anti-symmetric hypermultiplets, which could result in an anomaly-free spectrum. Since this procedure is precisely the starting point of the alternate approach that we will discuss momentarily, we simply state that there are no such $\fku(1)$s in $\hat{\fkh} = \fku(2) \oplus \fku(3)$, and so we conclude that the $\fksu(2)$ fundamental flavor symmetry is the unique subalgebra of the full flavor symmetry that can be gauged without altering the tensor or hypermultiplet spectrum; indeed, \cref{eq:su22su2fund} is precisely the automatic enhancement identified in \cite{Raghuram:2020vxm}.

Let us now see how we would reach this same conclusion by instead considering arbitrary $\fku(1)$ subalgebras of the flavor symmetry. We solve the system of equations \labelcref{eq:6dAC} for gauge algebra $\fksu(22) \oplus \fku(1)$ in terms of the multiplicities of the representations $\bm{1}_0$, $\bm{22}_{n + 22 f_1}$, $\bm{22}_{n + 22 f_2}$, $\bm{231}_{2 n + 22 a_1}$, $\bm{231}_{2 n + 22 a_2}$, $\bm{231}_{2 n + 22 a_3}$, $\bm{483}_0$, $\bm{1}_{22}$, $\bm{1}_{44}$, where we leave the integers $-10 < n \le 11$, $f_i$, and $a_i$ unspecified. We have chosen to normalize the $\fku(1)$ charges such that all charges will be integral regardless of the presence or absence of quotients in the global structure of the gauge group; as a result, two hypermultiplets transforming in the same $\fksu(22)$ representation must have $\fku(1)$ charges that differ by a multiple of $22$. The variable $n$ allows for offsets related to these possible quotients, while the variables $f_i, a_i$ allow the various hypermultiplets to have distinct charges. The offset of $2 n$ for the anti-symmetrics is required for consistency with the offset $n$ of the fundamentals.
We then look for solutions to the anomaly cancellation equations for which $x_{\bm{1}_{22}}, x_{\bm{1}_{44}} \le 0$, all other multiplicities are non-negative, and all multiplicities are integral. As discussed in \cref{sec:approach}, we can use the number of uncharged singlets in \cref{eq:spectrum_su22} to constrain the magnitudes of $f_i, a_i$ in our search for solutions. The resulting spectra with non-trivial $\fku(1)$ are all of the form
    \begin{equation}
        x_{\bm{1}_0} = 22\,, \quad x_{\bm{22}_{11 m}} = x_{\bm{22}_{-11 m}} = 1\,, \quad x_{\bm{231}_0} = 3\,, \quad x_{\bm{1}_{22 m}} = -2
    \end{equation}
for $m \in \bbZ$, with all other multiplicities zero. Here, we see the negative number of charged singlets indicating that the $\fku(1)$ factor we have gauged must be embedded in a gauged non-abelian subalgebra of $\hat{\fkh}$ in order to be anomaly-free. Specifically, there are two charged singlets in the branching rule of the $\fksu(2)$ adjoint to its Cartan $\fku(1)$ subalgebra, and so this is an indicator that the appropriate anomaly-free subalgebra of $\hat{\fkh}$ is in fact $\fksu(2)$; the fact that the $\fksu(22)$ fundamentals are charged under this $\fku(1)$ while the anti-symmetrics are not indicates that this is the fundamental flavor $\fksu(2)$, and the appropriate anomaly-free enhancement is the one given in \cref{eq:su22su2fund}. As mentioned above, this enhancement is precisely the one found in \cite{Raghuram:2020vxm}, and so we see that the arguments here reproduce the results of the original Automatic Enhancement Conjecture.

For the remaining examples, we will use only the shortcut approach in order to reduce the necessary computations. Next we consider the model with gauge symmetry $\fksu(20)$, $T = 0$, and hypermultiplet spectrum
    \begin{equation}
        x_{\bm{1}} = 23\,, \quad x_{\bm{20}} = 4\,, \quad x_{\bm{190}} = 3\,, \quad x_{\bm{399}} = 0\,,
    \end{equation}
which solves the anomaly cancellation conditions with $b = 1$. The flavor symmetry of this model is $\fku(4) \oplus \fku(3)$. We add an additional $\fku(1)$ gauge factor and solve the anomaly cancellation conditions \labelcref{eq:6dAC} with $b_{\fksu(20)} = 1$, allowing for potentially distinct $\fku(1)$ charges for each fundamental and anti-symmetric hypermultiplet. The spectra with non-trivial $\fku(1)$ and satisfying the requisite positivity conditions are all of the form
    \begin{equation}
        x_{\bm{1}_0} = 25\,, \quad x_{\bm{20}_{10 m}} = x_{\bm{20}_{-10 m}} = 2\,, \quad x_{\bm{190}_0} = 3\,, \quad x_{\bm{1}_{20 m}} = -2
    \end{equation}
for $m \in \bbZ$. As before, the negative number of charged singlets along with the charge assignments indicate that the appropriate anomaly-free subalgebra of $\hat{\fkh}$ is the $\fksu(2)$ fundamental flavor symmetry algebra, so the enhancement is to $\fksu(20) \oplus \fksu(2)$ with spectrum
    \begin{equation}
        x_{(\bm{1}, \bm{1})} = 25\,, \quad x_{(\bm{20}, \bm{1})} = 0\,, \quad x_{(\bm{190}, \bm{1})} = 3\,, \quad x_{(\bm{399}, \bm{1})} = 0\,, \quad x_{(\bm{20}, \bm{2})} = 2\,,
    \end{equation}
This again matches the enhancement found in \cite{Raghuram:2020vxm}.

Continuing with the cases discussed in Section~4.3 of \cite{Raghuram:2020vxm}, we consider the model with gauge symmetry $\fksu(19)$, $T = 0$, and hypermultiplet spectrum
    \begin{equation}
        x_{\bm{1}} = 26\,, \quad x_{\bm{19}} = 5\,, \quad x_{\bm{171}} = 3\,, \quad x_{\bm{360}} = 0\,,
    \end{equation}
which solves the anomaly cancellation conditions with $b = 1$. The flavor symmetry of this model is $\fku(5) \oplus \fku(3)$. Adding an additional $\fku(1)$ gauge factor and solving the anomaly cancellation conditions \labelcref{eq:6dAC} with $b_{\fksu(19)} = 1$, we find that all spectra with non-trivial $\fku(1)$ and satisfying the requisite positivity conditions are of the form
    \begin{equation}
        x_{\bm{1}_0} = 25\,, \quad x_{\bm{19}_{9 m}} = 5\,, \quad x_{\bm{171}_{-1 m}} = 3
    \end{equation}
for $m \in \bbZ$. We see in this case that there is no negative number of charged singlets, and so this $\fku(1)$ is not embedded in a non-abelian anomaly-free subalgebra of the flavor symmetry. Furthermore, both the fundamental and anti-symmetric representations are charged under this $\fku(1)$; this $\fku(1)$ is apparently a combination of $\fku(1)$s from within both factors of the $\fku(5) \oplus \fku(3)$ flavor symmetry. We can carry out the process again, adding another $\fku(1)$ factor to verify that this enhancement is maximal, and we find that indeed it is. This reproduces the enhancement of \cite{Raghuram:2020vxm}.

Finally, we consider the model with gauge symmetry $\fksu(18)$, $T = 0$, and hypermultiplet spectrum
    \begin{equation}
        x_{\bm{1}} = 30\,, \quad x_{\bm{18}} = 6\,, \quad x_{\bm{153}} = 3\,, \quad x_{\bm{323}} = 0\,,
    \end{equation}
which solves the anomaly cancellation conditions with $b = 1$. The flavor symmetry of this model is $\fku(6) \oplus \fku(3)$. We add an additional $\fku(1)$ gauge factor and solve the anomaly cancellation conditions \labelcref{eq:6dAC} with $b_{\fksu(18)} = 1$,
the spectra with non-trivial $\fku(1)$ and satisfying the requisite positivity conditions are all of the form
    \begin{equation}
        x_{\bm{1}_0} = 30\,, \quad x_{\bm{18}_{9 m}} = x_{\bm{18}_{-9 m}} = 3\,, \quad x_{\bm{153}_0} = 3
    \end{equation}
for $m \in \bbZ$. We again see that the $\fku(1)$ is maximal in this case, but this time it is a embedded solely in the $\fku(6)$ fundamental flavor symmetry. Adding another $\fku(1)$ factor and carrying out the process again, we can verify that this enhancement is indeed maximal, and it again matches the enhancement of \cite{Raghuram:2020vxm}.

\subsubsection{\texorpdfstring{$\fksu(3)$}{su(3)} gauge algebra with \texorpdfstring{$T = 0$}{T=0}}
\label{sec:su3_example}

In the case of $\fkg = \fksu(3)$, $T = 0$, we encounter our first example where the shortcut approach indicates that there are potentially multiple distinct maximal anomaly-free subalgebras of $\hat{\fkh}$. For $b_{\fksu(3)} = 6$, there is an enhancement of the gauge algebra by $\fku(1)^2$, as expected from \cite{Raghuram:2020vxm}. The potential issue appears when $b_{\fksu(3)} = 7$, with $M=42 \times {\bm{3}}$, which was seen in \cite{Raghuram:2020vxm} to exhibit enhancement to gauge group $G' = [\SU(3) \times \SU(3)] / \bbZ_3$ (see \cref{sec:1-form_anomalies} for further discussion of the global structure of the gauge group). If we solve the anomaly cancellation conditions for $\fksu(3) \oplus \fku(1)$ and impose the relevant constraints, we find four types of spectra with non-trivial $\fku(1)$ (after removing cases that amount to rescaling the charges):
    \begin{equation}
        \label{eq:su3threespectra}
        \begin{alignedat}{2}
            b_{\fku(1)} = 12\colon &&& \qquad x_{\bm{1}_0} = 40\,, \quad x_{\bm{3}_1} = 28\,, \quad x_{\bm{3}_{-2}} = 14\,, \quad x_{\bm{8}_0} = 15\,, \quad x_{\bm{1}_3} = -4\,,  \\
            b_{\fku(1)} = 36\colon &&& \quad
            \left\{\begin{gathered}
                 x_{\bm{1}_0} = 42\,, \quad x_{\bm{3}_0} = x_{\bm{3}_3} = x_{\bm{3}_{-3}} = 14\,, \quad x_{\bm{8}_0} = 15\,, \\
                 x_{\bm{1}_3} = -4\,, \quad x_{\bm{1}_6} = -2\,,
            \end{gathered}\right. \\
            b_{\fku(1)} = 54\colon &&& \qquad x_{\bm{1}_0} = 54\,, \quad x_{\bm{3}_3} = x_{\bm{3}_{-3}} = 21\,, \quad x_{\bm{8}_0} = 15\,, \quad x_{\bm{1}_3} = -18\\
            b_{\fku(1)} = 54\colon &&& \quad
            \left\{\begin{gathered}
                x_{\bm{1}_0} = 45\,, \quad x_{\bm{3}_6} = 1\,, \quad x_{\bm{3}_3} = 18\,, \quad x_{\bm{3}_0} = 3\,, \quad x_{\bm{3}_{-3}} = 20\,, \\
                x_{\bm{8}_0} = 15\,, \quad x_{\bm{1}_3} = -6\,, \quad x_{\bm{1}_6} = -3\,.
            \end{gathered}\right.
        \end{alignedat}
    \end{equation}
The first two spectra correspond to different $\fku(1)$ subalgebras of $\fkh = \fksu(3)$, and match with the expected $\fksu(3) \oplus \fksu(3)$ multiplicities of
    \begin{equation}
        x_{(\bm{1}, \bm{1})} = 43\,, \quad x_{(\bm{8}, \bm{1})} = 15\,, \quad x_{(\bm{3}, \bm{3})} = 14\,,
    \end{equation}
with anomaly coefficient $b_\fkh = 2$. The multiplicities and $\fku(1)$ charges of the flavor-$\fksu(3)$ fundamentals in these cases correspond to the branching rules of the bifundamental representation for these two embeddings, while the negative numbers of charged singlets correspond exactly to the numbers expected from the branching rules for the adjoint representation. The third and fourth spectra in \cref{eq:su3threespectra}, however, cannot correspond to an embedding of the $\fku(1)$ factor into an $\fksu(3)$, and thus potentially indicate the existence of a separate maximal anomaly-free subgroup of the flavor symmetry group. The final spectrum can immediately be ruled out from the distribution of charges of its $\fksu(3)$ fundamental representations, while the third spectrum requires more attention. In order to determine if this spectrum actually admits an embedding into a larger algebra, we must explicitly check the other possible non-abelian subalgebras of the $\fku(42)$ flavor symmetry. Note that the enhancement cannot be by multiple non-abelian factors, as if both factors had non-trivial anomaly coefficients $b$, then there would be a non-zero number of hypermultiplets charged under the bifundamental representation for these two factors, which cannot be the case for a subalgebra of a flavor symmetry. Carrying out this explicit check, we find that the third spectrum in \cref{eq:su3threespectra} does not correspond to an actual maximal subalgebra, and $\fksu(3)$ is the unique maximal subalgebra of the flavor symmetry that can be gauged.

This issue does not arise for $b_{\fksu(3)} = 8$; in this case, only the expected spectra corresponding to the enhanced gauge group $G' = [\SU(3) \times \SU(3)] / \bbZ_3$ found in \cite{Raghuram:2020vxm} appear.

\subsubsection{Other cases}

In addition to the examples worked out here, we have also checked all of the relevant enhancements from Sections~4.2 and 4.4\footnote{For more on these examples, see \cref{sec:e-string}.} of \cite{Raghuram:2020vxm}, and again find agreement with the enhancements presented there. These include cases where $T = 1$, which, as mentioned in \cref{sec:approach}, could potentially exhibit multiple allowable enhancements. Nevertheless, we again only find a unique gauge-able subalgebra $\fkh \subset \hat{\fkh}$ in all of these cases.

\subsection{Gauge group topology from gauged 1-form symmetries}
\label{sec:1-form_anomalies}

Another observation based on F-theory models, named ``Massless Charge Sufficiency Conjecture'' \cite{Morrison:2021wuv,Raghuram:2020vxm}, is that the absence of massless hypermultiplets transforming non-trivially under $\cZ$---a subgroup of the center $Z(\widetilde{G})$ of the simply-connected cover $\widetilde{G}$ for the non-abelian gauge algebra $\fkg$---automatically ``enhances'' the gauge group to $G = \widetilde{G}/\cZ$.
The description as an enhancement becomes literal, if one interprets the non-simply connected gauge group as the result of having gauged a $\cZ$ 1-form global symmetry of the $\fkg$ gauge theory \cite{Gaiotto:2014kfa}.
Then, cases that satisfy the Massless Charge Sufficiency Conjecture can be explained similarly as above, namely as quantum gravity gauging an anomaly-free global symmetry.

The anomaly in this case is a mixed anomaly between the 1-form center symmetry and the large gauge transformations of the (anti-)self-dual tensors originating in the Green--Schwarz term in the effective action  \cite{Apruzzi:2020zot}, which can be quantified as follows.
Let the gauge algebra $\fkg = \bigoplus_i \fkg_i$ have the simply-connected cover $\widetilde{G} = \prod_i \widetilde{G}_i$ with center $Z(\widetilde{G}) = \prod_i Z(\widetilde{G}_i) \cong \prod_i \bbZ_{n_i}$.\footnote{We are ignoring here the more subtle case $Z(\text{Spin}(4n)) = \bbZ_2 \times \bbZ_2$, see \cite{Apruzzi:2020zot} for details.}
Then, there is an obstruction to turning on a background field for---and, therefore, also to gauging---a subgroup $\cZ = \bbZ_m \subset Z(\widetilde{G})$ generated by $(k_i) \in \prod_i \bbZ_{n_i}$, if there exists a tensor with charge vector $b$ for which
\begin{equation}\label{eq:1-form_anomaly}
    \sum_i b \cdot b_i \, \alpha_{i} \, k_i^2 \notin \bbZ \, .
\end{equation}
This is a non-trivial constraint, since the numbers $\alpha_i$ associated to each gauge factor $\fkg_i$ with anomaly coefficient $b_i$ are in general fractional; for $\fkg_i = \fksu(n_i)$, we have $\alpha_i = \frac{n_i-1}{2n_i}$ \cite{Kapustin:2014gua} (see \cite{Cordova:2019uob} for values for other Lie algebras).

As an example, consider a $T=0$ theory containing a $\fkg = \fksu(2)$ gauge sector with anomaly coefficient $b_\fkg = 12$.
This is free of (0-form) gauge symmetries with only adjoint hypermultiplets, which are invariant under the center $\cZ = \bbZ_2$.
For any other tensor with charge $b$, the anomaly above, $b \cdot b_\fkg \, \alpha_\fkg = 3b \in \bbZ$, is always trivial.
Assuming that there are no other obstructions to turning on a $\cZ$ 1-form symmetry background, it would constitute a global symmetry, which must be gauged in a gravitational theory.
This gauging leads to an $\SO(3)$ gauge group, which is also reflected in the geometry of the F-theory realization \cite{Morrison:2021wuv,Raghuram:2020vxm}.
Another $T=0$ example is a $\fkg = \fksu(24)$ gauge algebra with $b_\fkg = 1$, which has 2-index anti-symmetric hypermultiplets.
This preserves a $\bbZ_2 \subset \bbZ_{24} = Z(\widetilde{G})$, which is generated by $k = 12 \mod 24 \in \bbZ_{24}$.
Since $\alpha_\fkg \, k^2 = \frac{23}{48} \, 12^2 = 69 \in \bbZ$, this center subgroup is also anomaly-free; again, the corresponding 1-form symmetry is automatically gauged in F-theory \cite{Morrison:2011mb} in the gauge group $G = \SU(24) / \bbZ_2$.

Moreover, the logic also applies to automatic enhancements $\fkg' = \fkg \oplus \fkh$ discussed above, where there is a resulting non-trivial global group structure $G' = [\widetilde{G} \times \widetilde{H}] / \cZ$ involving the enhanced flavor symmetry $\widetilde{H}$.
From the perspective of the original $\fkg$ gauge theory, this non-trivial global structure arises from a $\cZ$-twisted gauge bundle that is compensated by a corresponding $\cZ$-twist in the flavor bundle (see, e.g., \cite{Cohen:1983sd,Cherman:2017tey,Shimizu:2017asf,Gaiotto:2017tne}).
In particular, it means that there is the option to turn on a background 1-form $\cZ \subset Z(\widetilde{G})$ gauge field that is ``locked'' to an analogous background for the flavor symmetry.
In 6d $\cN=(1,0)$ theories, such a background can have a similar obstruction as above \cite{Apruzzi:2020zot}; since $\fkh$ is gauged through automatic enhancement, this anomaly boils down to including the anomaly coefficient $b_\fkh$ in the sum in \cref{eq:1-form_anomaly}.
The model discussed in \cref{sec:su3_example} provides an example of this sort.
There, the enhanced gauge algebra is $\fkg' = \fksu(3)_\fkg \oplus \fksu(3)_\fkh$, with $b_\fkg = 7$ and $b_\fkh = 2$, and the hypermultiplet spectrum contains only bifundamental matter, so a diagonal $\cZ = \bbZ_3$ generated by $(k_\fkg, k_\fkh) = (1,1) \in \bbZ_3 \times \bbZ_3 = Z(\SU(3) \times \SU(3))$ center is left unbroken.
Since this is a $T=0$ model, the only constraint comes from the self-dual tensor: $\alpha_{\fksu(3)} (b_\fkg k_\fkg^2 + b_\fkh k_\fkh^2) = \frac13 (7 + 2) \in \bbZ$.
So there is indeed no obstruction for the $\cZ$ 1-form symmetry, whose gauging leads to the gauge group $G' = [\SU(3) \times \SU(3)]/\bbZ_3$.

\subsection{Comment on the Massless Charge Sufficiency Conjecture}
\label{sec:comments}

Lastly, let us briefly comment on the exceptions to the Massless Charge Sufficiency Conjecture discussed in \cite{Morrison:2021wuv}, that is, cases where a simple gauge sector $\fkg$ has no massless non-adjoint representations, yet there is no non-trivial global gauge group, i.e., $G= \widetilde{G}$.
These fall into two categories, namely either when $\fkg$ is non-Higgsable, or when it is an $\cN=(1,1)$ gauge sector (i.e., its anomaly coefficient satisfies $b_\fkg \cdot b_\fkg = b_\fkg \cdot a = 0$), which is coupled to an $\cN=(1,0)$ supergravity system.

As standalone gauge theories, i.e., when decoupled from the rest, these indeed do not have the anomaly \labelcref{eq:1-form_anomaly} that would obstruct the center 1-form symmetry from being gauged; for the non-Higgsable clusters this was discussed in \cite{Apruzzi:2020zot}, and for the $\cN=(1,1)$ sectors, it follows from the absence of a Green--Schwarz term due to the properties of $b_\fkg$.
In F-theory, it turns out that the \emph{local} geometries describing such gauge theories do exhibit Mordell--Weil torsion, reflecting an unbroken 1-form center symmetry \cite{Aspinwall:1998xj,Mayrhofer:2014opa,Cvetic:2021sxm}.
For the non-Higgsable theories, whose 1-form symmetries were first analyzed in \cite{Morrison:2020ool}, we refer again to \cite{Apruzzi:2020zot} for the detailed relationship between Mordell--Weil torsion and absence of the anomaly \labelcref{eq:1-form_anomaly}.
For the $(1,1)$ sectors, the local base geometry is simply $T^2 \times \bbC$, where the torus is the curve associated to the anomaly coefficient $b_\fkg$.
Since the axio-dilaton is constant along the $T^2$, the elliptic fibration only changes non-trivially in the $\bbC$ direction, where there is a (non-compact) K3-surface with one singular fiber corresponding to the gauge algebra $\fkg$; equivalently, this is a trivial $T^2$ reduction of an 8d $\cN=1$ F-theory model with a $\fkg$ gauge symmetry.
The 8d gauge theory has an unbroken $Z(\widetilde{G})$ 1-form center symmetry \cite{Cvetic:2021sxm} (see also \cite{Morrison:2020ool,Albertini:2020mdx}), which therefore also survives after a trivial $T^2$ reduction.
Geometrically, it is reflected in the Mordell--Weil torsion preserved locally around the singular fiber in $\bbC$, which extends trivially to Mordell--Weil torsion over $\bbC \times T^2$.

We therefore see that the Massless Charge Sufficiency Conjecture does---if extended to non-gravitational theories---apply to the pure non-Higgsable clusters and $(1,1)$ gauge sectors: in the absence of massless hypermultiplets that break the 1-form center symmetry, the latter constitutes a global symmetry of the gauge theory.
From this perspective, it is clear that the violation of the conjecture, at least in the models considered in \cite{Morrison:2021wuv}, must arise from coupling to gravity.
Indeed, this can be easily understood from the mixed 1-form anomaly.
Namely, in these examples, there always exists at least one other dynamical tensor with charge $b$, such that \cref{eq:1-form_anomaly} is not an integer.
For the simplest gravitational models containing a non-Higgsable cluster, corresponding to F-theory on Hirzebruch surfaces $\bbF_n$, the tensor dual to the fibers of the ruling on $\bbF_n$ induces the anomaly.
Likewise, the simplest models with $(1,1)$ sectors have $T=9$, corresponding to F-theory on a rational elliptic dP$_9$ surface.
While the $(1,1)$ sectors live on the generic elliptic fibers of this surface, they are intersected by the sections of the fibration, which give rise to independent tensors whose large gauge transformations have a non-trivial anomaly \labelcref{eq:1-form_anomaly}.
One may wonder how these exceptions to the Massless Charge Sufficiency Conjecture satisfy the Completeness Hypothesis.
Clearly, since the gauge group is $\widetilde{G}$ rather than $\widetilde{G}/\cZ$, there must be massive states transforming non-trivially under $\cZ$.
As elaborated on in \cite{Apruzzi:2020zot}, these states are precisely provided by the excitations of the BPS strings that are dual to the tensor multiplets with a non-trivial anomaly \labelcref{eq:1-form_anomaly}.

We have not checked other examples apart from these ``simplest'' gravitational models that violate the Massless Charge Sufficiency Conjecture.
However, it is suggestive that such models must always contain tensor multiplets that give rise to the anomaly \labelcref{eq:1-form_anomaly}, by virtue of the necessary unimodularity of the tensor charge lattice.
It would be interesting to attempt such a proof, at least geometrically for models constructable in F-theory.

\section{Automatic enhancement enforced by BPS strings}\label{sec:strings}

We have seen that a large class of automatic enhancements is consistent with the gauging of an anomaly-free subalgebra $\fkh$ of the flavor symmetry of a $\fkg$ gauge theory.
However, the absence of anomalies is a necessary, but not sufficient condition for the appearance of the additional gauge sector $\fkh$.
That is, these arguments do not explain why the absence of additional gauge sectors would be \emph{inconsistent}.
Moreover, there are cases with hypermultiplets charged just under $\fkh$, although not enough to allow for a supersymmetric Higgsing; these cases do not allow for an interpretation of $\fkh$ as a flavor symmetry.

In this section, we analyze two classes of $T=1$ models without non-Higgsable clusters that exhibit automatic enhancements of this sort.
In these models, we can argue physically why certain gauge algebras $\fkg$ with anomaly coefficient $b_\fkg$ satisfying $b_\fkg \cdot b_1 = 1$ are inconsistent, where $b_1$ is the charge of the (unpaired) anti-self-dual tensor that necessarily has $b_1 \cdot b_1 = -1$ or $-2$.
The key ingredient is---by the Completeness Hypothesis---the existence of a dynamical string charged under $b_1$.
Under some mild assumptions, these strings can be identified with the E-string and the M-string (the string of an $\cN = (2, 0)$ $A_1$ SCFT) for $b_1 \cdot b_1 = -1$ and $b_1 \cdot b_1 = -2$, respectively.
Then, the inconsistency of models that are observed to automatically enhance is an incompatibility of the gauge symmetry $\fkg$ with the flavor symmetry of these SCFT strings.

\subsection{Automatic enhancement enforced by the E-string}\label{sec:e-string}

Let us first consider $T=1$ models with tensor pairing
\begin{equation}\label{eq:tensor_charge_lattice_F1}
    \Omega = \begin{pmatrix}
        0 & 1 \\
        1 & -1
    \end{pmatrix} \, ,
\end{equation}
where the basis is chosen such that the positivity cone is spanned by $(1,0)$ and $(0,1)$.
The gravitational anomaly coefficient is $a = (-3,-2)$.
In F-theory, such models are realized with the Hirzebruch surface $\bbF_1$ as the base manifold.

With such a tensor pairing, any 6d supergravity theory with gauge algebra $\fkg = \fksu(N>3)$ and anomaly coefficient $b_\fkg = (2,1)$ (with $b_\fkg \cdot b_\fkg = 3$ and $b_\fkg \cdot a = -5$) is anomaly-free with hypermultiplet spectrum $M = (40-3N) \times \bm{\mathrm{F}} \oplus 5 \times \bm{\mathrm{AS}}$, where $\bm{\mathrm{F}}$ (resp.~$\bm{\mathrm{AS}}$) denotes the fundamental (resp.~anti-symmetric) representation.
However, F-theory models with $N>8$ automatically enhance to $\fkg' \supset \fksu(N)$ \cite{Raghuram:2020vxm}, indicating the inconsistency of a bare $\fksu(N)$ gauge algebra in these cases.
Note that the enhancement patterns $\fkg' = \overline{\fkg} \oplus \fkh$ (where $\overline{\fkg}$ has anomaly coefficient $b_{\overline{\fkg}} = b_\fkg$) do not fit into the framework discussed previously, in that the additional gauge symmetry that appears in the F-theory constructions are not flavor symmetries of $(\fkg,M)$ (they have additional charged matter\footnote{
For the $N = 9$ case, the hypermultiplet spectrum of the enhanced model is $1\times(\bm{1}, \bm{2}) + 5\times(\bm{36}, \bm{1}) + 11\times(\bm{9}, \bm{1}) + 1\times(\bm{9}, \bm{2})$. For the $N = 10$ case, it is $5\times(\bm{45}, \bm{1}, \bm{1}) + 4\times(\bm{10}, \bm{1}, \bm{2}) + 1\times(\bm{10}, \bm{2}, \bm{1})$. For the $N = 11, 12$ cases, it is $5\times(\bm{66}, \bm{1}) + 1\times(\bm{12}, \bm{4})$.
In the $N=10$ case, the anomaly coefficient of the additional $\fksu(2)$ is $(2, 2)$.
}, which however does not allow for a supersymmetric Higgsing by itself):
\begin{equation}\label{eq:F1_enhancements}
{\arraycolsep = 10pt
\begin{array}{*{3}{c}} \toprule
    N & \overline{\fkg} & \fkh = \bigoplus_I \fkh_I \\ \midrule
    9 & \fksu(9) & \fksu(2) \\
    10 & \fksu(10) & \fksu(2) \oplus \fksu(2) \\
    11 & \fksu(12) & \fksp(2) \\
    12 & \fksu(12) & \fksp(2) \\ \bottomrule
\end{array}}
\end{equation}
The common feature of all these enhanced algebras is that the additional gauge factor $\fkh_1$ has anomaly coefficient $b_1 = (0,1)$.
This corresponds to an anti-self-dual tensor with $b_1 \cdot b_1 = -1$.
The validity of the Automatic Enhancement Conjecture would therefore imply that, for $N>8$, an unpaired tensor $b_1$, i.e., no gauge algebra with anomaly-coefficient $b_1$, is inconsistent.

To give a physical argument to this, observe that, by the Completeness Hypothesis, there must be a dual string charged under the tensor with charge $b_1$.
If $b_1$ is unpaired, there are compelling arguments \cite{Shimizu:2016lbw} based on anomaly considerations on the worldsheet of such a string that there must be an $E_8$ flavor symmetry in 6d; since it agrees with that of the E-string, it is generally believed that any such string must be an E-string.
When coupled to gravity, the worldsheet $\gE_8$ flavor symmetry is generically broken in the bulk.
A subalgebra $\fkg \subset \fke_8$ may be gauged if one can assign an anomaly coefficient $b_\fkg$, with $b_\fkg \cdot b_1 = 1$, to $\fkg$, provided all gauge and gravitational anomalies are cancelled (which may require hypermultiplets charged under $\fkg$).
However, it is clear that any such gauging can not ``exceed'' $\gE_8$, as it would be otherwise inconsistent with the worldsheet dynamics of the string.

This ``$\gE_8$-rule'' \cite{Witten:1996qb,Morrison:2012np,Heckman:2013pva,Johnson:2016qar} immediately explains why $\fkg = \fksu(N)$ with $N>9$ are inconsistent, as these $\fksu$ algebras are not contained in $\fke_8$.
By forcing such a gauge algebra with anomaly coefficient $b_\fkg = (2,1)$ in an F-theory construction \cite{Raghuram:2020vxm}, the observed automatic enhancement can be interpreted as the geometry attempting to rectify this inconsistency by pairing the string with charge $b_1 = (0,1)$ with a gauge symmetry $\fkh$.
In this case, the string is no longer an E-string, but rather an instanton string of $\fkh$, which is not subject to the ``$\gE_8$-rule''.
Note that this does not explain why, for $N=11$, the $\fksu(11)$ has to enhance to $\overline{\fkg} = \fksu(12)$, but it does rule out a bare $\fksu(11)$.

At first sight, the ``$\gE_8$-rule'' does not seem to explain the inconsistency of the model with $N=9$, since $\fksu(9) \subset \fke_8$.
However, it is important to realize that, at the group level, $\gE_8$ does not have an $\SU(9)$-subgroup, but rather $\SU(9)/\bbZ_3$.
To retain this non-trivial global structure, it means that any gauging of an $\fksu(9) \subset \fke_8$ subalgebra must have an unbroken $\bbZ_3 \subset \bbZ_9 = Z(\SU(9))$ 1-form symmetry.
In particular, this requirement demands the absence of any fundamental and 2-index anti-symmetric hypermultiplets.
Within the 6d SCFT landscape, the only possible gauging of this kind is $\fkg = \fksu(9)$ with anomaly coefficient $b_\fkg$ satisfying $b_\fkg \cdot b_\fkg = -2$ and $b_\fkg \cdot b_1 = 1$ (where $b_1$ is the tensor dual to the E-string), which has no massless hypermultiplets.
A supergravity model where an $\fksu(9) \subset \fke_8$ subalgebra of the E-string flavor symmetry is gauged can be constructed in F-theory on a dP$_9$ surface, i.e., a model with $T=9$, see \cref{app:su9_on_dP9}; in this geometric construction, the $\bbZ_3$ 1-form symmetry is manifestly gauged by the presence of Mordell--Weil torsion.\footnote{In fact, while in both the SCFT and the supergravity model, one would naively expect the full $\bbZ_9$ center to be preserved, the excitations of the E-string break it to an $\bbZ_3$ \cite{Bhardwaj:2020phs}, which can also be seen from the 1-form anomalies \cite{Apruzzi:2020zot}.
This is another exception to the Massless Charge Sufficiency Conjecture, as the $\fkg = \fksu(9)$ by itself forms an $\cN = (1,1)$ gauge sector, see \cref{sec:comments}.
}

Returning to the $T=1$ supergravity models with tensor pairing \labelcref{eq:tensor_charge_lattice_F1} and gauge symmetry $\fkg = \fksu(N)$ with anomaly coefficient $b_\fkg=(2,1)$, we now see that the inconsistency of $N=9$ model stems from the presence of the fundamental and anti-symmetric hypermultiplets, which are needed to cancel the gauge anomalies and which explicitly break the $\bbZ_3$ center symmetry.

\subsection{Automatic enhancement enforced by the M-string}

The simplest supergravity models that contain an M-string SCFT sector have a tensor pairing
\begin{equation}\label{eq:tensor_charge_lattice_F2}
    \Omega = \begin{pmatrix}
        0 & 1 \\
        1 & -2
    \end{pmatrix} \, ,
\end{equation}
where we have chosen the basis in which the positivity cone is spanned by $(1,0)$ and $(0,1)$, with the latter being the anti-self-dual tensor $b_1$ that has self-pairing $-2$.
The gravitational anomaly coefficient is $a = (-4,-2)$.
In F-theory, such models are realized with the Hirzebruch surface $\bbF_2$ as the base manifold.

We can easily verify that, just based on gauge and gravitational anomalies, a $\fkg = \fksu(N)$ model with $M = (48-4N) \times {\bf F} \oplus 6 \times {\bf AS}$ and anomaly coefficient $b_\fkg = (3,1)$ (satisfying $b_\fkg \cdot b_\fkg = 4, b \cdot a = -6, b_\fkg \cdot b_1 =1$) would be consistent for $4 \le N \le 9$; for $N=2,3$, the anomaly-free hypermultiplet spectrum would be $40 \times {\bf 2}$ and $42 \times {\bf 3}$, respectively.
However, if we attempt to engineer such models in F-theory, we see that there is automatic enhancement $\fkg' = \overline{\fkg} \oplus \fkh$ for $N \ge 3$, with an additional gauge factor $\fkh_1 \subset \fkh$ having anomaly coefficient $b_1$; see \cref{app:F-theory_on_F2} for details.

Again, a physical argument for the Automatic Enhancement Conjecture in this instance---i.e., why an unpaired tensor $b_1$ would be inconsistent for $N>2$---comes from the flavor symmetry of the string with charge $b_1$, which exists by the Completeness Hypothesis.
By a similar argument as for the E-string, one can show that the spacetime flavor symmetry is limited via worldsheet anomaly arguments to be $\fksu(2)$, which agrees with the flavor symmetry of what is known in the literature as the M-string \cite{Haghighat:2013gba}.\footnote{The name ``M-string'' originates from the M-theory realization as M2-branes ending on M5-branes, which have 6d $\cN=(2,0)$ worldvolume dynamics.}
Alternatively, such a string arises as the charged object of a 6d $\cN=(2,0)$ tensor multiplet (which also furnishes the tensor branch description of a $(2,0)$ $A_1$ SCFT).
When coupled to an $\cN = (1,0)$ background, an $\fksu(2)$ subalgebra of the $(2,0)$ R-symmetry may be interpreted as the flavor symmetry of the SCFT.\footnote{The $(2,0)$ R-symmetry is $\fksp(2) \cong \fkso(5)$, which in particular acts on the fermions in the $(2,0)$ tensor multiplet in the spinor representation $\bm{4}$ of $\fkso(5)$.
Under the branching $\fksp(2) \supset \fksu(2) \oplus \fksu(2)$, where one of the factors is identified with the flavor symmetry in the $\cN=(1,0)$ background, the decomposition $\bm{4} \rightarrow (\bm{2}, \bm{1}) \oplus (\bm{1}, \bm{2})$ yields fundamental degrees of freedom, thus indicating that the flavor group is $\SU(2)$.
This is confirmed in F-theory models, see \cref{app:F-theory_on_F2}.}
Therefore, any model with an $\fksu(N>2)$ symmetry on $b_\fkg$ with $b_\fkg \cdot b_1 =1$ can only give a globally consistent supergravity model if the tensor $b_1$ is paired with a non-trivial spacetime gauge symmetry itself.

\subsection{Discussion}
\label{sec:discussion}

It would be desirable to explain all instances of automatic enhancement using the consistency of BPS strings, as these provide, at least in the examples above, a definitive of proof why the lack of enhancement would be inconsistent.
In the case where these are strings of anti-self-dual tensors $b_1$, the distinction between consistent and inconsistent gauge symmetries $\fkg$ with anomaly coefficient $b_\fkg$ satisfying $b_\fkg \cdot b_1=1$ is analogous to the question of whether a $\fkg$ flavor symmetry of a 6d SCFT containing the tensor $b_1$ can be gauged (see \cite{Heckman:2018jxk} and references therein).
For supergravity models, the constraints are relaxed compared to SCFTs in the sense that the anomaly coefficient $b_\fkg$ of $\fkg$ can be a self-dual tensor, and $\fkg$ is allowed to have adjoint hypermultiplets (i.e., the genus of $b_\fkg$ can be non-zero).
As we have seen above in simple models, this still poses strong restrictions.
Moreover, even though we have not discussed them in detail, it is rather obvious that, in the presence of multiple anti-self-dual tensors $b_A$, there can be independent constraints on $\fkg$ with anomaly-coefficient $b_\fkg$ from each $b_A$ with $b_A \cdot b_\fkg \ne 0$.
By forcing the existence of the gauge algebra $\fkg$ in the corresponding F-theory model, one might then find a large, automatically enhanced algebra $\fkg' = \overline{\fkg} \oplus \bigoplus_A \fkh_A \oplus \dotsb$.
Examples of this type appear, e.g., in F-theory models on blow-ups of Hirzebruch surfaces, see \cite{Raghuram:2020vxm}.

The main challenge to extend this method to other cases of automatic enhancement, such as those discussed in \cref{sec:flavor_anomalies}, is that the strings in question are not SCFT strings, i.e., they are associated to tensors with $b_1 \cdot b_1 \ge 0$.
In the case $b_1 \cdot b_1=0$ and $b_1 \cdot a =-2$ (with $a$ the gravitational anomaly coefficient), the corresponding string may be identified with the heterotic string, which has either $\gE_8 \times \gE_8$ or $\operatorname{Spin}(32)/\bbZ_2$ flavor symmetry.
The latter contains an $\SU(N)$ subgroup for $N \le 15$, yet one observes automatic enhancements for $12 \le N \le 15$ in $\fksu(N)$ gauge symmetries with $b_\fksu$ such that $b_\fksu \cdot b_1 = 1$ \cite{Raghuram:2020vxm}.
Moreover, to the best of our knowledge, the maximal flavor symmetry of strings with $b_1 \cdot b_1 =1$ and $b_1 \cdot a = -2$, present in all $T=0$ supergravity models, has not been studied.

In $T=0$ models, one observes an additional type of enhancement pattern, $\fkg \to \fkg' \supset \fkg$ with a simple algebra $\fkg'$, discussed in \cite{Raghuram:2020vxm}, which does not fall into the classes discussed above.
Another instance of this type is the $N=11$ case in \cref{eq:F1_enhancements}, where $\fkg = \fksu(11)$ enhances to $\overline{\fkg} = \fksu(12)$, in addition to the appearance of $\fkh_1 = \fksp(2)$.
Presumably, this particular enhancement is related to the supergravity string present in this model, rather than the E-string.
Hence, a better understanding of the flavor symmetry of strings with $b_1 \cdot b_1 \ge 0$ will be necessary to scrutinize the physical explanation of the Automatic Enhancement Conjecture for these and more general cases.

\section*{Acknowledgments}

We thank Craig Lawrie for many valuable discussions and comments on the draft.
The work of MC and APT is supported in part by the DOE (HEP) Award DE-SC0013528. 
MC further acknowledges support by the Simons Foundation Collaboration grant \#724069 on ``Special Holonomy in Geometry, Analysis and Physics'', the Fay R.~and Eugene L.~Langberg Endowed Chair, and the Slovenian Research Agency (ARRS No. P1-0306).

\appendix

\section{Anomaly cancellation conditions}
\label{app:anomaly_conditions}

In this appendix, we collect the explicit anomaly cancellation conditions for a 6d $\cN = (1, 0)$ supergravity theory with tensor lattice $\Gamma$, pairing $\Omega(b_1, b_2) \equiv b_1 \cdot b_2$ for $b_1, b_2 \in \Gamma$, gauge algebra $\fkg = \bigoplus_i \fkg_i \oplus \bigoplus_\alpha \fku(1)_\alpha$, and matter \labelcref{eq:hyper_total_rep} \cite{Erler:1993zy,Green:1984bx,Sagnotti:1992qw,Park:2011ji}. 
The Green--Schwarz contributions cancel the 1-loop contributions from the various multiplets, as encoded in their respective anomaly polynomials \cite{Alvarez-Gaume:1983ihn}:
\begin{align}
    \begin{split}
        I^\text{grav}_8 & = -\frac{273}{5760} \left(\trace R^4 \frac54 + (\trace R^2)^2 \right) + \frac{9}{128} (\trace R^2)^2 \, , \\
        I^\text{tensor}_8 & = \frac{29}{5760} \left(\trace R^4 + \frac54 (\trace R^2)^2 \right) - \frac{1}{128} (\trace R^2)^2 \, , \\
        I_8^\text{vector}(\fkg_i) & = -\frac{\dim (\fkg_i)}{5760} \left(\trace R^4 + \frac54 (\trace R^2)^2 \right) - \frac{1}{24} \trace_{\bf adj} F_i^4 + \frac{1}{96} \trace_{\bf adj} F_i^2 \trace R^2 \, , \\
        I_8^\text{vector}(\fku(1)_\alpha) & = -\frac{1}{5760} \left( \trace R^4 + \frac54 (\trace R^2)^2 \right) \, ,
    \end{split} \label{eq:1-loop_contributions} \\
        I^\text{hyper}_8(\birrep) & = \frac{\dim (\birrep)}{5760} \left( \trace R^4 + \frac54 (\trace R^2)^2 \right) + \frac{1}{24} \trace_{\birrep} F^4 - \frac{1}{96} \trace_{\birrep} F^2 \trace R^2 \, ,
        \label{eq:hyper_1-loop}
\end{align}
where $\trace_{\birrep}$ denotes the trace in the representation $\birrep$.
In the hypermultiplet contribution, the traces over the combined field strength $F$ of the full gauge symmetry $\fkg$ can be disentangled into products of field strengths of the individual (non-)abelian factors, by applying recursively the formulae
\begin{align}\label{eq:decomposing_trace_non-ab}
    \trace_{\birrep}(F_\fkf^2) &= \dim(\birrep_1) \trace_{\birrep_2}(F_2^2) + \dim(\birrep_2) \trace_{\birrep_1}(F_1^2) \, , \\
    \trace_{\birrep}(F_\fkf^4) &= \dim(\birrep_1) \trace_{\birrep_2}(F_2^4) + 6 \trace_{\birrep_1}(F_1^2) \trace_{\birrep_2}(F_2^2) + \dim(\birrep_2) \trace_{\birrep_1}(F_1^4) \, ,
\end{align}
for $\fkf = \fkf_1 \oplus \fkf_2$ with $\fkf_i$ simple non-abelian and $\birrep = (\birrep_1 , \birrep_2)$, and
\begin{align}\label{eq:decomposing_trace_ab}
    \trace_{\birrep}(F_\fkf^2) &= \trace_{\birrep_\text{n}} (F_\text{n}^2) +  q_{\alpha} q_{\beta} F_{\alpha} F_{\beta} \, , \\
    \trace_{\birrep}(F_\fkf^4) &= \trace_{\birrep_\text{n}} (F_\text{n}^4) + 4 \trace_{\birrep_\text{n}}(F_\text{n}^3)  q_{\alpha} F_{\alpha} + 6 \trace_{\birrep_\text{n}}(F_\text{n}^2) q_{\alpha} q_{\beta} F_{\alpha} F_{\beta} + q_{\alpha} q_{\beta} q_{\gamma} q_{\delta} F_{\alpha} F_{\beta} F_{\gamma} F_{\delta} \, ,
\end{align}
for $\fkf = \fkf_\text{n} \oplus \bigoplus_{\alpha} \fku(1)_{\alpha}$ with $\fkf_\text{n}$ and $ \birrep = (\birrep_\text{n})_{(q_\alpha)}$. Here, summation over repeated Greek indices is implied.
Then, the vanishing of the anomaly polynomial $I_8^\text{GS} + I_8^\text{1-loop}$ is ensured term-wise if:
\begin{subequations}
    \label{eq:6dAC}
    \begin{alignat}{2}
        \trace(R^4)\colon && 0 &= H - V + 29 T - 273 \, , \label{eq:nonabelACgrav} \\
        \trace(R^2)^2\colon && a \cdot a &= 9 - T\,, \label{eq:nonabelACa} \\
        \trace(F_i^2) \trace(R^2)\colon && a \cdot b_i &= -\frac{1}{6} \lambda_i \left(\sum_r d_i^{(r)} A_{\irrep{\fkg_i}{r}} - A_{\fkg_i}\right)\,, \label{eq:nonabelACA} \\
        \trace(F_i^4)\colon && 0 &= \sum_r d_i^{(r)} B_{\irrep{\fkg_i}{r}} - B_{\fkg_i}\,, \label{eq:nonabelACB} \\
        \trace(F_i^2)^2\colon && b_i \cdot b_i &= \frac{1}{3} \lambda_i^2 \left(\sum_r d_i^{(r)} C_{\irrep{\fkg_i}{r}} - C_{\fkg_i}\right)\,, \label{eq:nonabelACC} \\
        \trace(F_i^2) \trace(F_j^2)\colon && b_i \cdot b_j &= \lambda_i \lambda_j \sum_r d_{i,j}^{(r)} A_{\irrep{\fkg_i}{r}} A_{\irrep{\fkg_j}{r}}\,, \quad i \ne j \,, \label{eq:nonabelACmix} \\
        \trace(R^2) F_\alpha F_\beta\colon && a \cdot b_{\alpha \beta} &= -\frac{1}{6} \sum_r d^{(r)} \charge{\alpha}{r} \charge{\beta}{r}\,, \label{eq:abelACsqr} \\
        \trace(F_i^3) F_\alpha\colon && 0 &= \sum_r d_i^{(r)} E_{\irrep{\fkg_i}{r}} \charge{\alpha}{r}\,, \label{eq:abelACE} \\
        \trace(F_i^2)F_\alpha F_\beta\colon && b_i \cdot b_{\alpha \beta} &= \lambda_i \sum_r d_i^{(r)} A_{\irrep{i}{r}} \charge{\alpha}{r} \charge{\beta}{r}\,, \label{eq:abelACA} \\
        F_\alpha F_\beta F_\gamma F_\delta\colon && \quad b_{\alpha \beta} \cdot b_{\gamma \delta} + b_{\alpha \gamma} \cdot b_{\beta \delta} + b_{\alpha \delta} \cdot b_{\beta \gamma} &= \sum_r d^{(r)} \charge{\alpha}{r} \charge{\beta}{r} \charge{\gamma}{r} \charge{\delta}{r}\,, \label{eq:abelACquar}
    \end{alignat}
\end{subequations}
where $d^{(r)} = x_r \prod_i \dim(\irrep{\fkg_j}{r})$, $d_i^{(r)} = x_r \prod_{j \ne i} \dim(\irrep{\fkg_j}{r})$, and $d_{i,j}^{(r)} = x_r \prod_{k \ne i, j} \dim (\irrep{\fkg_k}{r})$. Here, $H$, $V$, and $T$ are, respectively, the numbers of hypermultiplets, vector multiplets, and tensor multiplets in the theory. The quantities $A_{\birrep}, B_{\birrep}, C_{\birrep}, E_{\birrep}$ are group theoretic coefficients defined by
    \begin{equation}
        \trace_{\birrep} F^2 = A_{\birrep} \trace F^2\,, \quad \trace_{\birrep} F^4 = B_{\birrep} \trace F^4 + C_{\birrep} \left(\trace F^2\right)^2\,, \quad \trace_{\birrep} F^3 = E_{\birrep} \trace F^3\,,
    \end{equation}
with $\trace_{\birrep}$ the trace in representation $\birrep$ and $\trace$ the trace in the fundamental (or defining) representation. We use $A_{\fkg_i}, B_{\fkg_i}, C_{\fkg_i}, E_{\fkg_i}$ to denote these coefficients for the adjoint representation of $\fkg_i$.
The values of the coefficients $\lambda_i$ are as follows:
\begin{equation}
    \begin{array}{*{10}{c}} \toprule
                    \fkg & \fksu(N) & \fkso(N) & \fksp(N) & \fke_6 & \fke_7 & \fke_8 & \fkf_4 & \fkg_2 \\ \midrule
            \lambda & 1     & 2     & 1     & 6     & 12    & 60    & 6     & 2 \\ \bottomrule
        \end{array}
\end{equation}

\section{E-string with gauged \texorpdfstring{$\SU(9)/\bbZ_3$}{SU(9)/Z3} flavor symmetry in F-theory}
\label{app:su9_on_dP9}

In this appendix, we give a concrete F-theory example where an $\SU(9)/\bbZ_3$ gauge group is located next to a $(-1)$-curve.
In particular, this provides a simple example of a ``non-Tate'' tuning of the $\fksu(9)$ subalgebra of the $\fke_8$ flavor symmetry of an E-string, which was not possible with Tate's algorithm \cite{Johnson:2016qar}.

Such an F-theory geometry is an elliptic fibration over a $B=\operatorname{dP}_9$ surface, given by the Weierstrass model $y^2 = x^3 + f x + g$ with
\begin{equation}
    f = \frac12 a_1 b_1^3 - \frac{1}{48}a_1^4 \, , \quad g = \frac14 b_1^6 + \frac{1}{864} a_1^6 - \frac{1}{24} a_1^3 b_1^3 \, ,
\end{equation}
which is just the generic Weierstrass model with $\bbZ_3$ Mordell--Weil torsion \cite{Aspinwall:1998xj}, specialized to $a_3 = b_1^3$.
From the discriminant, $\Delta = 4f^3 + 27g^2 = \frac{1}{16} b_1^9 (3b_1 - a_1) (a_1^2 + 3a_1 b_1 + 9b_1^2)$, we see that there is an $I^\text{s}_9$ fiber over $\{b_1=0\}$ (splitness follows from $\left.9g/2f\right|_{\{b_1=0\}} = -a_1^2/4$ being a perfect square), which corresponds to an $\fksu(9)$ gauge algebra.

Note that the divisor classes $[\{a_1 =0\}] = [\{b_1=0\}]$ are just the anti-canonical class $-K$ of $B = \operatorname{dP}_9$, which coincides with the fiber class of the rational elliptic surface $\operatorname{dP}_9 \rightarrow \bbP^1$.
Then, $\{a_1 = 0\}$ and $\{b_1 = 0\}$, for generic choices of sections of $\mathcal{O}_B(-K)$, correspond to two distinct, smooth, and irreducible elliptic fibers of this fibration.
This means that, apart from the $\fksu(9)$ on $\{b_1=0\}$, the F-theory model described by the Weierstrass model above has no other non-abelian gauge factor.
Moreover, since generic fibers do not intersect, there are no codimension-two singularity enhancements, i.e., no massless matter other than the single adjoint hypermultiplet, associated to the genus of the elliptic curve $\{b_1=0\}$ on which the $\fksu(9)$ lives.

So far, we have only made manifest the $\SU(9)/\bbZ_3$ gauge group on $\{b_1=0\}$, with the $\bbZ_3$ quotient coming from the Mordell--Weil group.
To see that this gauges part of the $\gE_8$ flavor symmetry of an E-string, recall that a $\operatorname{dP}_9$ surface also has sections of the elliptic fibration $B \rightarrow \bbP^1$, which themselves are genus 0 curves with self-intersection $(-1)$, i.e., give rise to E-strings via wrapped D3-branes.
Since a section $S$ intersects any smooth fiber exactly once, in particular also $\{b_1=0\}$, we see that the $\SU(9) / \bbZ_3$ indeed can be identified with the gauged flavor symmetry of the E-string on $S$.
Note that the $\operatorname{dP}_9$ surface actually has 9 independent (in homology) sections, so that the $\SU(9)/\bbZ_3$ on $\{b_1=0\}$ is in fact the ``diagonal'' flavor symmetry of all 9 (independent) E-string sectors.

\section{Automatic enhancement for F-theory on \texorpdfstring{$\bbF_2$}{F2}}
\label{app:F-theory_on_F2}

In this appendix, we describe the F-theory models on a $B = \bbF_2$ base that undergo automatic enhancement.
To begin with, we recall that a Hirzebruch surface is a $\bbP^1$-bundle over a base $\bbP^1$.
We introduce homogeneous coordinates $[s:t]$ for the fiber $\bbP^1$, and $[u:v]$ for the base $\bbP^1$, and denote by $[x]$ the divisor class of the codimension-one variety $\{x=0\}$.
Then, the second homology lattice of $\bbF_2$, which is identified with the tensor lattice, is generated by $[s]$ and $[v]$, with linear relations $[s] = [t]$, $[u] = [v] + 2[s]$, and intersection numbers $[s]^2 = 0, [s] \cdot [v] = 1, [v]^2 = -2$.
The canonical class $K$, which is identified with the gravitational anomaly coefficient $a$, is $K = - (4[s] + 2[v])$.

The geometric origin of automatic enhancement can be attributed to generic representatives of certain homology classes becoming reducible \cite{Raghuram:2020vxm}.
Such a representative can be expressed in terms of the vanishing locus $\{p=0\}$ of a homogeneous polynomial of the appropriate multi-degree, which becomes reducible if any such polynomial factorizes.
For a representative of an effective divisor class $n [s] + m [v]$, with $n,m \ge 0$, the homogeneous polynomials are
\begin{equation}
    p = v^{m} \, q_{n} + v^{m -1} u \, q_{n-2} + ... + v^{m-k} u^k \, q_{n-2k} + ... + u^m q_{n-2m} \, ,
\end{equation}
where $q_{d}$ are homogeneous polynomials of degree $d$ in $(s,t)$.
If $d<0$, then $q_d \equiv 0$, which means that a generic polynomial of multi-degree $n[s]+m[v]$ factorizes, 
\begin{equation}\label{eq:factorization_condition_on_F2}
    p = v^{m- \lfloor n/2 \rfloor} \, \tilde{p} \,, \quad \text{if and only if} \quad n < 2m \,.
\end{equation}
Note that a generic representative of any multiple of the anti-canonical divisor class $-K = 4[s] + 2[v]$ is irreducible.

The relevant supergravity models considered in \cref{sec:strings} have $\fksu(N)$ gauge symmetry with anomaly coefficient $b_\fkg = (3,1) = 3[s]+[v]$, which corresponds to a curve class $[C]$ with $[C]^2 = 4, [C] \cdot K = -6$, and, hence, a generic representative is an irreducible rational curve.
In F-theory, engineering an $\fksu(N)$ on a generic curve $C = \{\sigma=0\}$ in this homology class requires a Weierstrass model $y^2 = x^3 + f\,x+g$ to take the following forms \cite{Morrison:2011mb,Katz:2011qp}
\begin{equation}
    \begin{aligned}
        N & = 2k \colon & f &= -\frac13 \Phi^2 + f_k \,\sigma^k \, , \quad g = \frac{2}{27} \Phi^3 - \frac13 \Phi \, f_k \, \sigma^k + g_N \, \sigma^N \, , \\
        N & = 2k+1 \colon & f &= -\frac13 \Phi^2 + \frac12 \phi_0 \psi_k \sigma^k + f_{k+1} \sigma^{k+1} \, , \\
        & & g &= \frac{2}{27} \Phi^3 - \frac{\Phi}{6}(\phi_0 \psi_k + 2f_{k+1}\sigma) \sigma^k + \frac14 \psi_k^2 \sigma^{2k} + g_N \sigma^N \, ,
    \end{aligned}
\end{equation}
with $\Phi = \frac14 \phi_0^2 + \phi_1 \sigma$.

Focusing on $N=2k$ first, observe that the polynomials $f_k$ and $g_N$ have divisor classes
\begin{equation}
\begin{split}
    [f_k] & = [f] - k[\sigma] = -4K - k[\sigma] = (16-3k)[s] + (8-k)[v] \, , \\
    [g_N] & = [g] - N[\sigma] = -6K - N[\sigma] = (24 - 6k)[s] + (12 - 2k)[v] \, .
\end{split}
\end{equation}
By \cref{eq:factorization_condition_on_F2}, we have $f_k = v^{\lceil k/2 \rceil} \tilde{f}_k$ and $g_N =v^k \tilde{g}_N$ for $k\ge 1$, where $\tilde{f}_k$ and $\tilde{g}_N$ are irreducible.
Note that, since $[\phi_0] = -K$ does not factorize, $\Phi^2$ and $\Phi^3$ do not contain any non-trivial power of $v$ as common factor.
This means that $f$ and $g$ remain generically irreducible (as long as $[f_k]$ and $[g_N]$ are effective).
Then, it is straightforward to check that the discriminant of the Weierstrass model factorizes as
\begin{equation}
    4f^3 + 27g^2 = \sigma^N \left(4 f_k^3 \sigma^k + 27 g_N^2 \sigma^N - 18 f_k g_N \Phi \sigma^k - f_k^2 \Phi^2 + 4g_N \Phi^3\right)
    = \sigma^N v^{k} (\dotsb)  \, .
\end{equation}
Note that for $k=1$, i.e., an $\fksu(2)$ on $\{\sigma=0\}$, this factorization gives rise to an $I_1$ singularity over $\{v=0\}$, which does not give rise to any gauge symmetry.
For larger $k$, we find an $I^\text{s}_k$ fiber, which gives rise to an $\fksu(k)$ gauge symmetry over $\{v=0\}$.
Note that the splitness here comes about because $[\Phi] = -2K$ has trivial intersection number with $[v]$, so $\Phi$ does not vanish anywhere on $\{v=0\}$, hence $\left.9g/2f\right|_{\{v=0\}}$ is a perfect square.

In the odd ($N=2k+1$) cases, the relevant divisor classes are
\begin{equation}
\begin{split}
    [\psi_k] & = [f] - [\phi_0] - k[\sigma] = -3K - k[\sigma] = (12-3k)[s] + (6 - k)[v] \, , \\
    [f_{k+1}] & = [f] - (k+1)[\sigma] = -4K - (k+1)[\sigma] = (13-3k)[s] + (7-k) [v] \, , \\
    [g_N] & = [g] - N[\sigma] = -6K - N[\sigma] = (21 - 6k) [s] + (11 - 2k) [v] \, ,
\end{split}
\end{equation}
which again all factorize by \cref{eq:factorization_condition_on_F2}:
\begin{equation}
    \psi_k = v^{\lceil k/2 \rceil} \tilde{\psi}_k \, , \quad f_{k+1} = v^{\lfloor \frac{k+2}{2} \rfloor} \tilde{f}_{k+1} \, , \quad g_N = v^{k+1} \tilde{g}_N \, .
\end{equation}
By carefully distinguishing between even and odd $k$, we find a factorization of the discriminant:
\begin{equation}
    4f^3 + 27g^2 = \begin{cases}
        \sigma^N v^{k} \, (\dotsb) & k \, \text{ even} \, , \\
        \sigma^N v^{k+1} \, (\dotsb) & k \, \text{ odd} \, .
        \end{cases}
\end{equation}
This gives rise to either an $\fksu(k)$ (for even $k$) or an $\fksu(k+1)$ (for odd $k$) gauge symmetry on $\{v=0\}$.

In summary, we see that F-theory constructions on $\bbF_2$ of supergravity models with $\fkg = \fksu(N)$ and anomaly coefficient $b_\fkg = (3,1) = 3[s]+[v]$ always force a non-trivial gauge symmetry $\fkh$ along the $(-2)$-curve $\{v=0\}$ when $N>2$, indicating an automatic enhancement with anomaly coefficient $[v] \equiv b_\fkh= (0,1)$.
For $N=2$, where there is no enhancement, the presence of 40 fundamental hypermultiplets confirms that the gauge group is $G = \SU(2)$.

\bibliographystyle{JHEP}
\bibliography{refs.bib}

\end{document}